\documentclass[3p]{elsarticle}

\usepackage{comment}
\usepackage{amssymb}
\usepackage{amsmath}
\usepackage[ruled,vlined,linesnumbered]{algorithm2e}
\usepackage{algorithmic}
\usepackage{tabularx,ragged2e,booktabs,caption}
\usepackage{mathtools}
\usepackage{graphicx}
\usepackage{balance}
\usepackage{subcaption}
\usepackage{relsize}

\newtheorem{defn}{Definition}

\DeclarePairedDelimiter\ceil{\lceil}{\rceil}

\journal{Journal of \LaTeX\ Templates}

\biboptions{authoryear}

\begin{document}
\begin{frontmatter}

\title{Lighting Two Candles With One Flame: An Unaided Human Identification Protocol With Security Beyond Conventional Limit
 \tnoteref{mytitlenote}}

\author[]{Nilesh Chakraborty\corref{cor1}}
  \ead{nilesh.pcs13@iitp.ac.in}
\author[]{Samrat Mondal}
  
  \cortext[cor1]{Corresponding author}
  \address{Department of Computer Science and Engineering}
  \address{Indian Institute of Technology, Patna}
  \address{Bihar - 801103, India}

\begin{abstract}
Designing an efficient protocol for avoiding the threat of recording based attack in presence of a powerful eavesdropper remains a challenge for more than two decades. During authentication, the absence of any secure link between the prover and verifier makes things even more vulnerable as, after observing a threshold challenge-response pair, users' secret may easily get derived due to information leakage. Existing literatures only present new methodologies with ensuring superior aspects over previous ones, while ignoring the aspects on which their proposed schemes cope poorly. Unsurprisingly, most of them are far from satisfactory $-$ either are found far from usable or lack of security features.

 To overcome this issue, we first introduce the concept of ``leakage control" which puts a bar on the natural information leakage rate and greatly helps in increasing both the usability and security standards. Not just prevention, but also, by introducing the threat detection strategy (based on the concept of \textit{honeyword}), our scheme ``lights two candles". It not only eliminates the long terms security and usability conflict under the practical scenario, but along with threat detection from client side, it is capable of protecting the secret at the server side under the distributed framework, and thus, guaranteeing security beyond the conventional limit.   
\end{abstract}


\end{frontmatter}


\section{Introduction}
\label{intro}
Password base authentication is one of the simplest form of authentication as it reduces the human effort to a great extent during the identity verification \citep{bonneau2012quest}. Being usable, this factor of authentication has been challenged under different kind of threats over the times \citep{rainbow-tab} \citep{dictionary-attack} \citep{attack-password-01} \citep{attack-password-02} \citep{attack-password-03} \citep{targeted-attack}. Though most of these threats have been successfully handled \citep{sec-password-02} \citep{sec-password-01} \citep{sauth}, there are a few, particularly those which involve human intelligence factor, are continuously challenging researchers in developing some efficient algorithm to tackle the breaches. Recording attack is one such security threat (on client side) which has severe impact on the password based authentication \citep{leakage-resilient-challenge}.\\

\textbf{Threat model:} Let a genuine user and adversary be denoted by $\mathcal{H}$ and $\mathcal{A}$, respectively. During registration, it is always assumed that $\mathcal{H}$ successfully submits her login credentials to a remote machine (identified as $\mathcal{M}$) in a private environment. At the time of authentication, $\mathcal{H}$ sends her login information to $\mathcal{M}$ by using a login terminal. Throughout an authentication session, with the help of some recording devices (e.g., conceal camera), $\mathcal{A}$ may record the complete login information submitted by $\mathcal{H}$. Later, she may use that captured information to impersonate $\mathcal{H}$. This kind of threat is known as observation attack, more preciously recording attack, on the password based authentication. In this context, one important aspect is \textbf{number of observations} that $\mathcal{A}$ can make, and based on this, the following two categories are suggested in \citep{type-adversary}.

\begin{itemize}
\item \textit{Type 1:} Video captures the entire authentication process only once.
\item \textit{Type 2:} Video captures the entire authentication process more than once (denoted as $\textit{r}_\text{max}$ times).\\
\end{itemize}

It is believed that while nature of the threat involving \textit{Type 1} adversary is opportunistic, \textit{Type 2} adversary engages herself in a planned attack \citep{pitfalls-recording-attack}. Though nature of $\mathcal{A}$ can really be either of these in practice, but some salient researches in this direction select value of $\textit{r}_\text{max}$ no greater than $3$ \citep{record-less3s-01} \citep{record-less3s-02} \citep{record-less3s-03} \citep{record-less3s-04} \citep{record-less3s-05}. These references will be particularly helpful during security analysis of our proposal.\\

\textbf{Defense strategy:} Defense to this threat model mainly relies on a basic principle of challenge ($\mathcal{C}$) response ($\mathcal{R}$) protocol. During login, $\mathcal{M}$ sends $\mathcal{C}$ (or a puzzle) to $\mathcal{H}$. Based on her original password (identified as $\mathcal{P}$), $\mathcal{H}$ then derives $\mathcal{R}$ with respect to $\mathcal{C}$. Thus, we may roughly present $\mathcal{R}$ in the form of \textit{f}($\mathcal{P}$, $\mathcal{C}$). It is important to note here that generated $\mathcal{C}$ by  $\mathcal{M}$ varies in each authentication session, because of which $\mathcal{R}$ also changes. Therefore passive key entry (i.e., submitting $\mathcal{R}$ instead of $\mathcal{P}$) by $\mathcal{H}$ refrains $\mathcal{A}$ to get the original password.\\  

In this paper, we revisit the working principle of recording attack resilient passive key entry methods that do not require any secure auxiliary channel, and study their security aspects. There is no denial of the fact that no usage of auxiliary link mitigates the risk of different side channel attacks to a great extent \citep{side-channel-timing-attack}. Also, it helps $\mathcal{H}$ to login without being dependent on any additional hardware. But as $\mathcal{C}$ is overtly communicated, therefore, due to information leakage \citep{core-paper}, these methods cannot withstand the exposure of too many authentication sessions. \\

\textbf{Motivations and Contributions:} After performing an exhaustive literature survey, we have found that schemes that do not rely on a secure channel to address the considered threat model, end up being unusable for most of the users \citep{strong-ssa-hopper} \citep{strong-ssa-pas} \citep{strong-ssa-s3pas} \citep{strong-ssa-sp}. This in turn justifies the proposed claim\footnote{To achieve both security and usability against considered threat model, a scheme must rely on a certain secure channel.} by Yan et al. in \citep{core-paper}. 

Also, we have found that to address the password leakage from a compromised server (indicating to server side threat), recently proposed threshold password-only authentication schemes \citep{address-server-threat-distributed} (store password information over multiple servers, and thus no alliance of servers upto a certain threshold can learn anything about the secret) are inherently unable to cope with the recording based attack at the client side. Therefore, in order to address all the aforementioned security aspects, we have made the following major contributions in this paper. 

\begin{itemize}
\item \textbf{Contribution 1:} We propose a ``two passwords based protocol", namely TPP $-$ in which a second password is introduced as the ``second line of defense", that provides security against recording based attack. The usage of two passwords here not only increases the overall security at the client side, but also provides a distributed security framework to address the threat of password leakage from a compromised server.

\item \textbf{Contribution 2:} We introduce a novel idea of controlling the information leakage rate explicitly. Explicit control of information leakage helps in defeating $\mathcal{A}$ for more number of authentication sessions.

\item \textbf{Contribution 3:} We show that ``second line of defense" not only prevents, but also detects the attackers' activity to a certain extent to ``light two candles with one flame".

\item \textbf{Contribution 4:}  Along with the experimental survey, we also provide a theoretical analysis to measure the usability standard of the proposed approach. The usability study infers that proposed scheme significantly reduces the workload and ensures almost same usability standard as of the legacy password authentication.\\ 
\end{itemize}

\textbf{Roadmap:} The rest of the paper is organized as follows. Section \ref{prelim} gives some basic information that will be helpful to understand the proposed idea. Along with the concept of leakage control and threat detection, Section \ref{proposed} introduces the proposed methodology. Section \ref{storage} then deals with the storage mechanism of users' password to fit into our proposal.  A detailed security and usability analysis of the method is preformed then in Section \ref{sec-ana} and Section \ref{use-ana}, respectively. Followed by this, Section \ref{comp} presents a detailed comparative analysis of our scheme with the existing usable protocols in this direction. Finally, Section \ref{conclusion} concludes on outcome of our contribution.

\section{Preliminaries}
\label{prelim}
This section deals with some relevant information and ideas that will be helpful to understand the foundation of the proposed protocol. The discussion here will be directed through the following topics.

\begin{itemize}
\item[1.] Target user.
\item[2.] Interaction between $\mathcal{H}$ and $\mathcal{M}$.
\item[3.] Attack analysis and information leakage.
\item[4.] Basics behind the proposed thought.  
\end{itemize}
 
\subsection{Target user} 
\label{pre-1}
Likewise existing state of arts in \citep{kwon2014} \citep{ColorPass}, the design of the proposed TPP protocol here is based on the colors. We mainly use $4$ colors to materialize the proposed concept. However, as shown in \citep{kwon2014}, for the color blind people ($4.5\%$ of the total population), all $4$ utilized colors in our protocol can be replaced by $4$ symbols (like black, white, strips and dots). Therefore, like legacy password, anybody can use our scheme, except the blind people.

\subsection{Interaction between $\mathcal{H}$ and $\mathcal{M}$}
\label{pre-2}
During registration, along with a username, $\mathcal{H}$ selects two passwords. From a password space of all printable characters, the first password ($\mathcal{P}_1$) can be of any length (longer than a threshold value) decided by $\mathcal{M}$. According to our design, being of the length $4$, the second password ($\mathcal{P}_2$) belongs to a password space of $64$ characters containing $\{$A, B, ..., Z, a, b, ..., z, 0,1, ..., 9, *, $\#\}$. Both these passwords allow multiple occurrences of any character.

In TPP, during login, along with the username, $\mathcal{H}$ transcribes $\mathcal{P}_1$ through the legacy user interface (UI) as shown in Figure \ref{legacy-UI}.

\begin{figure*}[!ht]
\centering
\begin{tabular}{|c c|}
\hline
\textbf{Enter username} & alice\\
\textbf{Enter first password} & *********\\
\hline
\end{tabular} 
\caption{Legacy UI accepts username and $\mathcal{P}_1$} 
\label{legacy-UI}
\end{figure*}

After providing the aforementioned information, $\mathcal{H}$ indirectly inputs a secret bit from $\mathcal{P}_2$ by using
the proposed login interface detailed in Section \ref{pro-3}.
 In a nutshell, TPP integrates the legacy UI with the proposed idea in Section \ref{pro-3} for accepting $\mathcal{H}'$s password information in two phases.

\subsection{Attack analysis and information leakage}
\label{pre-3}
As discussed earlier, the core of any user authentication system that does not involve any auxiliary hidden link can be explained by a function of the form 

\begin{equation}
\textit{f} : \mathcal{P} \times \mathcal{C} \longrightarrow \mathcal{R}
\end{equation}
    
The absence of the hidden link together with considered threat model implies that the value of $\mathcal{C}$ in any phase of authentication will not just be shared between $\mathcal{H}$ and $\mathcal{M}$, but be fully accessed by $\mathcal{A}$. Like any other authentication system, we assume that no physical occlusion of user input is involved so that $\mathcal{R}$ is fully disclosed to everyone. Thus, having full access to both $\mathcal{R}$ and $\mathcal{C}$, a powerful eavesdropper can derive a set of possible secrets containing the original $\mathcal{P}$.  The attack principle can be realized in the form of the following function 

\begin{equation}
\textit{g} : \mathcal{R} \times \mathcal{C} \longrightarrow \mathcal{S}
\end{equation}
where the set $\mathcal{S}$ contains all possible $\mathcal{H}'$s secrets including the actual $\mathcal{P}$.\\

Let $\mathcal{S}_i$ be the derived $\mathcal{S}$ by $\mathcal{A}$ after recording the $i^{\text{th}}$ authentication session. As values of $\mathcal{R}$ and $\mathcal{C}$ vary in each session, consequently, $\mathcal{S}$ also differs. Thus, for two different authentication sessions, $i$ and $j$,  derived $\mathcal{S}_i$ and  $\mathcal{S}_j$ become different. Therefore from the recorded footages of these two authentication sessions, $\mathcal{A}$ can perform an intersection between $\mathcal{S}_i$ and $\mathcal{S}_j$ to reduce the probable candidate elements. 

Hence, the leaked information or information leakage after recording $k^{\text{th}}$ ($> 0$) authentication session can be presented in the form of the following equation.      

\begin{equation}
 \text{Leaked Information} =
  \begin{cases}
    \mathcal{S}_k       &  \text{if } k \text{= 1}\\
    \bigcap_{k=1}^i \mathcal{S}_k & \text{if } k \text{= 1, 2, ..., i}\\
  \end{cases}
\end{equation}

Above equation suggests that leaked information after each authentication session shrinks the search space in favour of $\mathcal{A}$. Let the entropy \citep{entropy} of $\mathcal{P}$ be $\mathcal{E}$. We denote the entropy of information leakage, resulting in secret space shrinking in each session, as $\triangle \mathcal{E}$ (inevitably $\triangle \mathcal{E} > 0$).  Therefore, after being used for $\lceil \mathcal{E}/\triangle\mathcal{E} \rceil$ authentication sessions, $\mathcal{P}$ gets exposed to $\mathcal{A}$. Under this situation, after utilizing it for $\lceil \mathcal{E}/\triangle\mathcal{E} \rceil$ authentication sessions, $\mathcal{H}$ needs to change her $\mathcal{P}$.\\

\textbf{Note 1: Premature attack $-$} It is important to note here that after recording multiple sessions, $\mathcal{A}$ may derive a small list of  probable secrets. In future, if $\mathcal{A}$ does not get a chance to record any further authentication session, then also she may try to login by using those secrets one by one, and eventually discovers the actual $\mathcal{P}$. Though this is a serious concern, but remains unaddressed in the existing literatures. 

\subsection{Basics behind the proposed thought}
\label{pre-4}
Discussion from the previous section reveals few important facts which may help in increasing the session resiliency of a defense mechanism against the recording attack. To defeat $\mathcal{A}$ for more number of sessions, $\mathcal{M}$ may

\begin{itemize}
\item expand of the secret space to increase $\mathcal{E}$.
\item minimize the leakage rate. 
\item make use of both.
\end{itemize} 

To fulfil the first option, $\mathcal{H}$ requires to remember more information. But due to limited capacity of human mind this is difficult to achieve. Therefore very little work follow this direction \citep{strong-ssa-sp} \citep{longpassword}. Methods that satisfy second criterion, often make login process complex for $\mathcal{H}$ \citep{strong-ssa-hopper} \citep{acn}. Complex login procedure threats practicality of an approach to be used by the common people. It is quite intuitive that third alternative, combining the first two options, will not be able to maintain any kind of balance between the usability and security aspects.

Our study reveals the fact that easy to use methodologies suffer from high information leakage rate \citep{kwon2014} \citep{weak-ssa-bw}. As a consequence, sometimes $\mathcal{A}$ becomes able to recover the secret from a single session recording only. Thus, if leakage rate can be controlled then with the same usability standard, session resiliency of a scheme can be hiked. In this paper, we have tried to achieve this. Due to leakage control, let entropy of the saved information in each session be $\triangle\mathcal{T}$ (inevitably $\triangle\mathcal{E} > \triangle\mathcal{T}$ and $\triangle\mathcal{T} > 0$). Therefore, explicit control of information leakage improves the session resiliency as the following equation stands.

\begin{equation}
   \ceil*{\dfrac{\mathcal{E}}{\triangle\mathcal{E}-\triangle\mathcal{T}}}   >   \ceil*{\dfrac{\mathcal{E}}{\triangle\mathcal{E}}}
\end{equation} 

Control over $\triangle\mathcal{T}$ helps in managing the shrinking rate of the secret space. Next, we define the \textit{shrinking factor} which relates the number of recorded sessions to the likelihood of  $\mathcal{A}$ successfully authenticating herself.

\begin{defn}
\textbf{Shrinking factor } is the rate of secret space shrinking after $\mathcal{A}$ records each authentication session.  
\end{defn}

If  $\mathcal{A}$ derives $\mathcal{S}_i$ after recording $i^{\text{th}}$ ($\geq 2$) authentication session then \textit{shrinking factor} (\textit{sf}) can be expressed as

\begin{equation}
\textit{sf} = \dfrac{\bigg|\bigcap_{k=1}^{i-1} \mathcal{S}_k\bigg|}{\bigg|\bigcap_{k=1}^i \mathcal{S}_k\bigg|} 
\end{equation}

where $|\mathcal{S}|$ denotes cardinality of the set $\mathcal{S}$.\\

\textbf{Addressing the premature attack:} For addressing the issue mentioned in the \textit{Note 1}, the following strategy  to \textit{distinguish between $\mathcal{H}$ and $\mathcal{A}$} has been adopted. 

\begin{itemize}
\item $\mathcal{M}$  stores the leaked information in the database after successful login attempt by $\mathcal{H}$ in an authentication session.

\item In the immediate next session, $\mathcal{M}$ will construct few groups by using those obtained leaked information from the previous session. The formed groups will be mutually exclusive while one of them will definitely contain the original $\mathcal{P}$.

\item If submitted $\mathcal{R}$ in this session corresponds to any other group except one, holding the original $\mathcal{P}$, then $\mathcal{M}$ will detect that $\mathcal{A}$ is trying to login by using the recording footage (leaked information) of the previous authentication session.

\item On detecting this attack, $\mathcal{M}$ acts according to the security policy set by the system administrator.
\end{itemize}

Detection strategy here is influenced by the concept of another threat detection mechanism, namely \textit{honeyword} (false password) based authentication technique \citep{fred-cohen} (ref. to appendix A), currently is being used in many domains \citep{honeypot-fake-session}.

\section{Proposed methodology}
\label{proposed}
As $\mathcal{P}_1$ is transcribed through a legacy UI, thus it does not require any further introduction. In this section, we mainly focus on how $\mathcal{H}$ makes use of the second password ($\mathcal{P}_2$) to defeat $\mathcal{A}$ from performing the recording attack. Contribution in this section unfolds in the following three phases. 

\begin{itemize}
\item The first phase introduces \textit{Basic Color Identification Protocol} (or BCIP) that (like any other method) leaks the information in an unrestricted way and provides no security for detecting the threat.

\item The second phase presents few basic principles for controlling the leakage rate and gives a direction towards the threat detection.

\item Lastly, on the basis of proposed principles, we have modified BCIP and shown that by restricting the leakage rate, the revised scheme is capable of detecting the security breach with enhanced session resiliency. The modified scheme has been named as \textit{Improved Color Identification Protocol} or ICIP.
\end{itemize}

\subsection{BCIP: Proposed strategy with convention}
\label{pro-1}
\textbf{Processing the elements of visual interface:} As mentioned earlier, in addition to two special symbols $*$ and $\#$, password space of $\mathcal{P}_2$ is comprising of all the alphabets (in both the cases) and the non-negative single digits from $0$ to $9$. To design the visual interface, $\mathcal{M}$ assigns all these characters to a $8 \times 8$ grid by following a specific order (e.g., alphabets first in the alphabetical order, the digits thereafter and the symbols at last). A convenient order helps $\mathcal{H}$ in locating her password character from the grid during login. As stated in Section \ref{pre-1}, we have used $4$ colors for coloring the grid in BCIP. The color assignment in BCIP is done  by obeying the following rules

\begin{itemize}
\item Allotment of the colors follows no specific order.
\item Each color appears on $16$ different cells.
\item Position of the colors changes in each round of an authentication session.\\
\end{itemize}

\textbf{Mediums for communicating $\mathcal{C}$:} During authentication, $\mathcal{M}$ first sends  $\mathcal{C}$ to $\mathcal{H}$. Human only have five traditional senses to obtain a stimulus: sight, hearing, touch, smell and taste. As long as an application cannot encode an information by generating last two stimuli, therefore $\mathcal{M}$ cannot make use of taste or smell to transfer any information.  As a result, sight, hearing and touch remain as possible candidates. Here, we explore the visual channel to transfer $\mathcal{C}$ overtly from $\mathcal{M}$ to $\mathcal{H}$. \\

\textbf{Interaction between $\mathcal{H}$ and $\mathcal{M}$:} After entering the first password ($\mathcal{P}_1$) information,   $\mathcal{M}$ sends an integer value (\textit{t}) between $1$ and $4$ (length of $\mathcal{P}_2$) to $\mathcal{H}$, by means of a visual signal. Starting from the first index, $\mathcal{H}$ then retrieves the character placed at $t^{\text{th}}$ index position of $\mathcal{P}_2$. Let the obtained character be denoted by $\eta$ ($\in \mathcal{P}_2$). A reference to $\eta$ is then used to generate $\mathcal{R}$ in each round of that authentication session. 

At each round, $\mathcal{M}$ randomly shuffles the allotments of the colors on the grid. To pass an authentication round, $\mathcal{H}$ needs to identify the color appeared at the grid's cell containing $\eta$. Let the identified color be $\eta^{\text{C}}$. Login interface of BCIP contains $4$ different color buttons which are used as the mode of $\mathcal{H}'$s interaction with the system. Each color button holds one of those $4$ colors that has been used to color the grid. $\mathcal{H}$ then presses $\eta^{\text{C}}$ button  to make $\mathcal{M}$ understand the identified color by her.\\

\begin{figure}[!htb]
\minipage{0.32\textwidth}
  \includegraphics[width=0.7\linewidth]{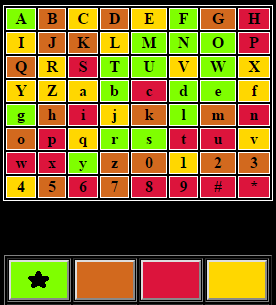}
  \subcaption{Round 1: $\mathcal{R}$ is Green} \label{fig:a}
\endminipage\hfill
\minipage{0.32\textwidth}
  \includegraphics[width=0.7\linewidth]{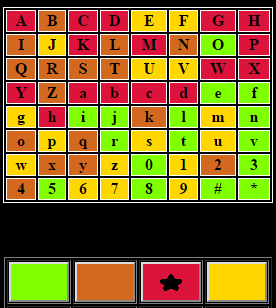}
  \subcaption{Round 2: $\mathcal{R}$ is Red} \label{fig:b}
\endminipage\hfill
\minipage{0.32\textwidth}%
  \includegraphics[width=0.7\linewidth]{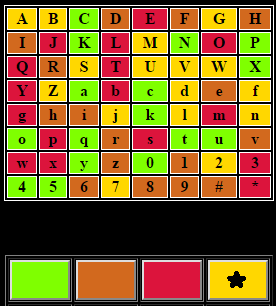}
  \subcaption{Round 3: $\mathcal{R}$ is Yellow} \label{fig:c}
\endminipage

\minipage{0.32\textwidth}
  \includegraphics[width=0.7\linewidth]{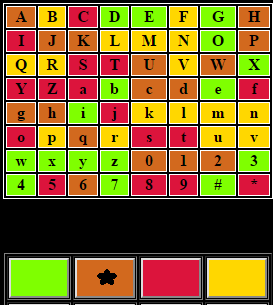}
 \subcaption{Round 4: $\mathcal{R}$ is Orange} \label{fig:d}
\endminipage\hfill
\minipage{0.32\textwidth}
  \includegraphics[width=0.7\linewidth]{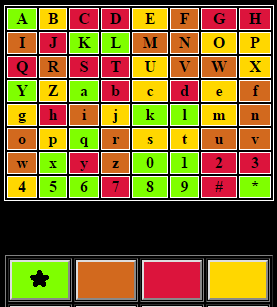}
  \subcaption{Round 5: $\mathcal{R}$ is Green} \label{fig:e}
\endminipage\hfill
\minipage{0.32\textwidth}%
  \includegraphics[width=0.7\linewidth]{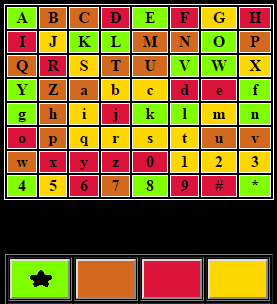}
  \subcaption{Round 6: $\mathcal{R}$ is Green} \label{fig:f}
\endminipage

\caption{\textbf{Login example in BCIP:} Above figure shows color responses by $\mathcal{H}$ for the password character ``\textit{A}" in each round of the authentication session. The color button hit by $\mathcal{H}$ in each round is marked by  a ``$\mathlarger{\mathlarger{*}}$" symbol.}  \label{nlc}

\end{figure}

\textbf{Login example:} Let the chosen $\mathcal{P}_2$ by $\mathcal{H}$ be \textit{S7Ay}. We also assume that $-$ like the 6 digit PIN entry method in \citep{6-digit-PIN}, each authentication session in BCIP is comprising of $6$ login rounds. Without loss of generality, if $\mathcal{H}$ receives $3$ as $\mathcal{C}$, then the selected password character by her would be \textit{A}. Thereafter in each subsequent round of that session, $\mathcal{H}$ will submit that color which will appear on the alphabet \textit{A} on the grid. Figure \ref{nlc} shows an instance of the login procedure in a session for the password character \textit{A}. We have used \textit{green}, \textit{orange}, \textit{red} and \textit{yellow} as $4$ different colors to design BCIP.\\

\textbf{Attack procedure and information Leakage in BCIP:}  From the recorded video, $\mathcal{A}$ first looks at the color button pressed by  $\mathcal{H}$ in each authentication round. Thereafter, $\mathcal{A}$ finds all those characters from the grid which are of the same color as of that color button. Hence, the response of $\mathcal{H}$ from the first round confuses $\mathcal{A}$ among $16$ possibilities. For the login example shown in Figure \ref{nlc}, leaked information after the first round  produces  $\mathcal{S}_\text{r1}$ containing $\{$A, F, M, N, O, T, U, W, b, d, e, g, l, r, s, y$\}$. Below we have shown the leaked information at the end of each round for the specific authentication session, illustrated in Figure \ref{nlc}.

\begin{itemize}
\item \textit{First round:} Leaked information $\mathcal{S}_\text{r1}$ = $\{$A, F, M, N, O, T, U, W, b, d, e, g, l, r, s, y$\}$.
\item \textit{Second round:} $\mathcal{S}_\text{r2}$ = $\{$A, C, D, G, H, K, M, P, W, X, Y, a, b, c, d, h$\}$. Leaked information $\mathcal{S}_\text{r1} \cap \mathcal{S}_\text{r2}$ = $\{$A, M, W, b, d$\}$.
\item \textit{Third round:} $\mathcal{S}_\text{r3}$ = $\{$A, B, G, M, S, U, V, W, Z, d, f, l, j, n, 2,7$\}$. Leaked information $\mathcal{S}_\text{r1} \cap \mathcal{S}_\text{r2} \cap \mathcal{S}_\text{r3}$ = $\{$A, M, W, d$\}$.   
\item \textit{Fourth round:} $\mathcal{S}_\text{r4}$ = $\{$A, H, J, K, P, U, W, c, d, g, h, q, 0, 1, 2, 6$\}$. Leaked information $\mathcal{S}_\text{r1} \cap \mathcal{S}_\text{r2} \cap \mathcal{S}_\text{r3} \cap \mathcal{S}_\text{r4}$ = $\{$A, W, d$\}$.
\item \textit{Fifth round:} $\mathcal{S}_\text{r5}$ = $\{$A, K, L, Y, a, k, l, q, x, 0, 1, 5, 6, 8, 9, *$\}$. Leaked information $\mathcal{S}_\text{r1} \cap \mathcal{S}_\text{r2} \cap \mathcal{S}_\text{r3} \cap \mathcal{S}_\text{r4}\cap \mathcal{S}_\text{r5}$ = $\{$A$\}$.

\end{itemize}
Though very simple to use, but above example shows that BCIP cannot even withstand the exposure of a single authentication session.\\

 All the password characters in $\mathcal{P}_2$ are independent of each other and $\mathcal{H}$ only uses a single bit from $\mathcal{P}_2$ to pass through the BCIP. Thus, along with estimating the leakage rate in BCIP, next we will analyze the security provided by a single password character against the recording attack.\\

\textbf{Estimating the leakage rate:} We state that after analyzing an authentication round, if $\mathcal{A}$ is confused among \textit{m} ($> 1$) possibilities then the next round has probability 1-P$_\text{Disclosure}$ of containing enough information to guard the secret of being leaked. Here P$_\text{Disclosure}$ is the probability associated with: in a round, none (inevitably except the original password character) of the {\textit{m-1}} probable elements from the previous round gets the same color as of the original password character. P$_\text{Disclosure}$ can be formalized by using the following equation. 

\begin{equation}
\text{P}_{\text{Disclosure}} = \prod_{k=1}^{m-1} \dfrac{(64-k)-(m-1)}{64-k} 
\label{eqnMSV}
\end{equation}
  
  It is quite obvious that after recording the $1^{\text{st}}$ round, $\mathcal{A}$ derives $m$ (here $16$) probable elements. This implies that in the next round, the color of the original password character must appear on at least one of the remaining $15$ characters from the first round to obfuscate $\mathcal{A}$. This yields the value of P$_\text{Disclosure}$ as $0.007$. Therefore, at the end of second round, without any leakage control mechanism, BCIP can guard the password character of $\mathcal{H}$ with the probability $0.993$. But as the value of $m$ goes low down the propagation of each login round, hence, after recording a complete session, chances of revealing the actual password character becomes significantly high.\\

\begin{figure}[!h]
\centering
\includegraphics[width=0.5\textwidth]{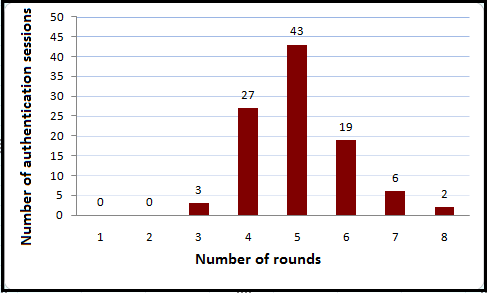}
\caption{On an observation of $100$ authentication sessions, above figure shows session resiliency of BCIP under the recording attack. }
\label{simulation}
\end{figure}

\textbf{Simulating the strength of BCIP against recording attack:} Figure \ref{simulation} shows the effect of information leakage on BCIP while experimented for a hundred authentication sessions. We found that in $92\%$ of scenarios, BCIP cannot even protect the secret for a single authentication session. For rest of the cases, the secret has been protected for two authentication sessions only.\\

\noindent\fbox{%
    \parbox{1.0\textwidth}{%
        \textbf{Discussion 1:} \textbf{Pitfalls of BCIP$-$} Proposed BCIP works on a simple idea of color identification to authenticate a genuine $\mathcal{H}$. Though the scheme is easy to use, but it is too weak to handle the recording attack. It is also obvious that from the submitted $\mathcal{R}$, BCIP cannot distinguish between a genuine user and the eavesdropper. 

    }%
}

\subsection{Basic Principles For Leakage Control And Threat Detection} 
\label{pro-2}   
For the leakage control and threat detection, $\mathcal{M}$ adopts the following strategy.

\begin{itemize}
\item Sets the value of \textit{shrinking factor} as \textit{sf} ($> 1$).
\item From the valid $\mathcal{R}$ in the first authentication session, $\mathcal{M}$ builds an initial group of $\mathcal{X}$ elements to confuse $\mathcal{A}$ among $\mathcal{X}$ probable candidates.
\item From the \textbf{next session onwards}, $\mathcal{M}$ does the following.

 \begin{itemize}
 \item Determines the cardinality of each subgroup as $\mathcal{Y} = \mathcal{X}/sf$.
 \item Builds $\ceil*{ \mathcal{X}/\mathcal{Y}} $ subgroups from those $\mathcal{X}$ elements of the previous session in such a 	   manner so that generated subgroups are mutually exclusive.
 \item Assigns a unique key (here color) to each subgroup in each round of the authentication session.
 \item Selects the subgroup containing the original password character, only after $\mathcal{H}$ successfully passes all the 		   authentication rounds. 
 \item In the database, replaces the group of $\mathcal{X}$ elements (obtained from the previous session) by the elements of 		   the selected subgroup.
 \item Updates the value of $\mathcal{X}$ as:  $\mathcal{X} \leftarrow \mathcal{Y}$.\\
 \end{itemize}
\end{itemize}

\textbf{Leakage control and threat detection:} From the first session, let $\mathcal{M}$ groups $\mathcal{X}$ ($> 1$) elements in such a manner so that each element from that group appears as a probable secret to $\mathcal{A}$. Therefore, after recording the first session, $\mathcal{X}$ elements create an illusion in $\mathcal{A}'$s mind. As recording attack threats security of a system only after $\mathcal{A}$ records atleast one authentication session, thus any threat detection strategy would not be able to differentiate between the submitted $\mathcal{R}$ by $\mathcal{H}$ and $\mathcal{A}$ in the first authentication session.

From the recorded footage of the first session, $\mathcal{A}$ remains confused among $\mathcal{X}$ elements. Now, with the goal of minimizing the probable secret space, $\mathcal{A}$ can proceed for the recording attack in the second session. To resist $\mathcal{A}$ in this session, $\mathcal{M}$ does the following trick. From the $\mathcal{X}$ elements in the first session, $\mathcal{M}$ creates few subgroups of cardinality $\mathcal{Y} = \mathcal{X}/sf$. 
 During assignment of a key (e.g., color) to each subgroup, $\mathcal{M}$ takes care of the following principles.  

\begin{itemize}
\item Each subgroup must get a unique key.
\item Because of the key, generated $\mathcal{R}$ must be unique with reference to the each subgroup.
\end{itemize}

Such an assignment of keys plays a dual role to resist and detect the recording attack.\\

\textbf{Leakage control and threat prevention:} Let the cardinality of each subgroup in a session be $\mathcal{Y}$ $(>1)$. From the recorded footage of the previous session, $\mathcal{A}$ will certainly be able to identify each subgroup . But deterministically, she would not be able to spot that subgroup containing the original password character. Now each subgroup holds a different key, which also makes an impact on $\mathcal{R}$. Therefore, after recording $\mathcal{R}$ by the genuine $\mathcal{H}$, though $\mathcal{A}$ can identify the proper subgroup holding the original password character, but remains confused among the $\mathcal{Y}$  probable candidates belonging to that subgroup. Thus, grouping of objects successfully controls the information leakage rate to defeat $\mathcal{A}$ for more number of authentication sessions.\\

\textbf{Threat detection:} After recording a session, if $\mathcal{A}$ tries to login in the next session then her first task would be to identify each subgroup made of those $\mathcal{X}$ elements from the previously recorded session. Identifying the subgroups would not reveal any clue about the original password character and hence, each subgroup will seem to be equally likely to $\mathcal{A}$. Let the number of subgroups be $\mathbb{G}^n$ $(> 1)$. This obfuscates $\mathcal{A}$ among $\mathbb{G}^n$ possibilities. Due to the key assignment property, each subgroup generates a unique sequence of responses. Therefore from the submitted $\mathcal{R}$, $\mathcal{M}$ will be able to identify the intended subgroup. As a result, if $\mathcal{A}$ chooses a wrong subgroup (not containing the original password character) and gives her responses accordingly, then $\mathcal{M}$ can detect the threat with the probability $(\mathbb{G}^n-1)/\mathbb{G}^n$. After detecting the $\mathcal{A}'$s activity, $\mathcal{M}$ may block that account depending on the security policy.\\

\noindent\fbox{%
    \parbox{1.0\textwidth}{%

\textbf{Discussion 2:} \textbf{Formalizing the threat detection and leakage control}$-$ With the change in value of $\mathcal{X}$ (updated as $\mathcal{X} \leftarrow \mathcal{X}/sf$ in each session), $\mathcal{M}$ follows the same strategy for leakage control and threat detection in the subsequent sessions. $\mathcal{A}$ gets the original password character only after recording a session where each subgroup holds a single element. 
This can be formalized as; grouping of possible response elements can resist and detect the attack for $1 + \lceil{ \text{log}^{\text{init}(\mathcal{X})}_{sf} }\rceil$ and $\lceil{ \text{log}^{\text{init}(\mathcal{X})}_{sf} }\rceil$ authentication sessions, respectively; where init($\mathcal{X}$) denotes the initial value of $\mathcal{X}$ after recording the very first authentication session. 
 }
}\\

Discussion $2$ shows that values of $\text{init}(\mathcal{X})$ and \textit{sf} play a big role in achieving the desired security objectives. Therefore, next we discuss the exact relationship between these two parameters.\\

\noindent\fbox{%
    \parbox{1.0\textwidth}{%

\textbf{Discussion 3:} \textbf{Formalizing the relationship between $\mathcal{X}$ and \textit{sf}}$-$ By setting a large initial value of $\mathcal{X}$, $\mathcal{M}$ can partially fulfil the criterion of defeating $\mathcal{A}$ for more number of authentication sessions. To earn more session resiliency, $\mathcal{M}$ should also set a small value of \textit{sf}. Therefore \textit{security against recording attack} (\textit{SR}) is influenced by $\mathcal{X}$ and \textit{sf} in the following manner

\begin{equation}
SR \propto \dfrac{\mathcal{X}}{sf}
\end{equation} 

In contrast, probability of threat detection  (Pr[\textit{TD}]) goes high  if $\mathcal{M}$ can create more number of subgroups in a session. To create more subgroups, values of  both  $\mathcal{X}$ and \textit{sf}  must be increased which can be understood in the form of the following equation.

\begin{equation}
\text{Pr}[\textit{TD}] \propto \mathcal{X} \cdot sf
\end{equation}

Therefore, due to contradictory nature of \textit{sf}, we must balance both these values (e.g., large value of $\mathcal{X}$ and moderate value of \textit{sf}) in such a manner so that both the threat detection and prevention strategies can complement each other.
 }
}

\subsection{ICIP: Revised strategy with the leakage control}
\label{pro-3}
Discussion $1$ from Section \ref{pro-1} infers that BCIP leaks too many information, and thus, most of the time cannot even protect the secret under a single session recording attack.  Like BCIP, each character of $\mathcal{P}_2$ in ICIP participates independently in an authentication session. Therefore, we will focus on a single password character to demonstrate the mechanisms of the leakage control and threat detection. 
Prior to elaborating the proposed leakage control technique, it is important to note here that BCIP and ICIP differ only in terms of \textit{processing the elements of visual interface}. The all other aspects (e.g., \textit{password space}, \textit{medium for communicating $\mathcal{C}$} etc.) remain same. \\

\textbf{Forming the initial group of init($\mathcal{X}$) elements:}  We have used $4$ colors (resemble to the keys in Section \ref{pro-2}) to color $64$ elements. Because of the uniform distribution of colors, each color gets assigned to $16$ elements. The elements that get the same color in the first round, create a group for that entire authentication session. Thus ICIP forms $4$ groups of cardinality $16$ in the first round and ensures that for remaining rounds in that session, all elements belonging to a group acquire the same color. If received $\mathcal{R}$ by $\mathcal{M}$ in each round matches with the color of the group containing the original password character then $\mathcal{M}$ validates the prover. After successful login operation by $\mathcal{H}$, the database stores all the elements of the group holding the actual secret bit.

Discussion $3$ shows that to make $\mathcal{A}'$s job difficult, selecting the proper values for init($\mathcal{X}$) (here $16$) and \textit{sf}  is absolutely important, and our next discussion (ref. to Discussion $4$) focuses on this.\\

\noindent\fbox{%
    \parbox{1.0\textwidth}{%

\textbf{Discussion 4:} \textbf{Choosing the balanced values for $\mathcal{X}$ and \textit{sf}}$-$ We have analyzed that if values of init($\mathcal{X}$) and \textit{sf} can be related in the form of the following equation, 
\begin{equation}
(\textit{sf})^{d} = \text{init}(\mathcal{X}) \quad \text{where } d \in \mathbb{Z}^+
\end{equation}
then the following facts stand

\begin{itemize}
\item Except the first, the detection probability of the recording attack in each session becomes $\frac{\textit{sf}-1}{\textit{sf}}$ (inevitably \textit{sf}$> 1$).

\item Including the first one, a method can defeat $\mathcal{A}$ for 1+  \text{log}$^{\text{init}(\mathcal{X})}_{sf}$ authentication sessions.
\end{itemize}

Thus for init($\mathcal{X}$) $= 16$, if we choose a moderate value of \textit{sf} as $4$, (as $4^2 = 16$) then ICIP performs reasonably well as each password character can hide the secret for $3$ authentication sessions and detect the recording attack with a probability $0.75$. 

 }
}\\

\begin{figure}[t!] 
\minipage{0.32\textwidth}
\includegraphics[width=0.7\linewidth]{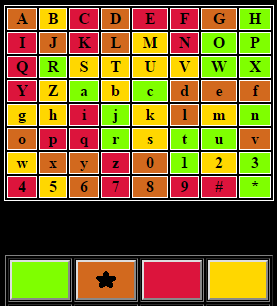}
\subcaption{Round 1: $\mathcal{R}$ is Orange} \label{fig:a}
\endminipage\hfill
\minipage{0.32\textwidth}
\includegraphics[width=0.7\linewidth]{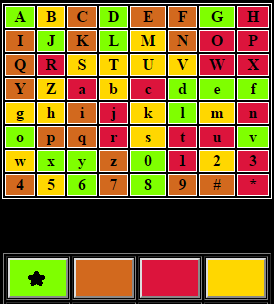}
\subcaption{Round 2: $\mathcal{R}$ is Green} \label{fig:b}
\endminipage\hfill
\minipage{0.32\textwidth}
\includegraphics[width=0.7\linewidth]{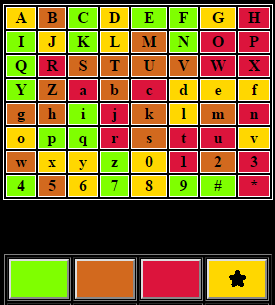}
\subcaption{Round 3: $\mathcal{R}$ is Yellow} \label{fig:c}
\endminipage

\minipage{0.32\textwidth}
\includegraphics[width=0.7\linewidth]{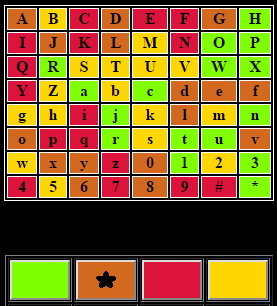}
\subcaption{Round 4: $\mathcal{R}$ is Orange} \label{fig:d}
\endminipage\hfill
\minipage{0.32\textwidth}
\includegraphics[width=0.7\linewidth]{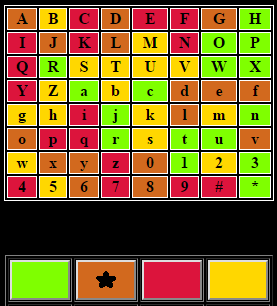}
\subcaption{Round 5: $\mathcal{R}$ is Orange} \label{fig:e}
\endminipage\hfill
\minipage{0.32\textwidth}
\includegraphics[width=0.7\linewidth]{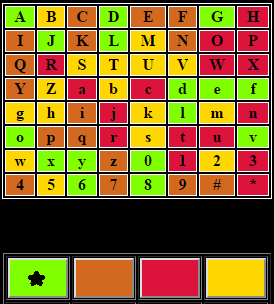}
\subcaption{Round 6: $\mathcal{R}$ is Green} \label{fig:f}
\endminipage

\caption{\textbf{Login example in ICIP:} Color responses submitted by $\mathcal{H}$ for the password character \textit{A} in each round of the \textbf{first authentication session}. The formed group here $\{$A, D, G, J, L, d, e, f, l, o, v, x, y, 0, 6, 8 $\}$ holds the original password character in each round. The color button hit by $\mathcal{H}$ in each round is marked by a ``$\mathlarger{\mathlarger{*}}$" symbol.}
  \label{fslc} 
\end{figure}

In Figure \ref{fslc}, we have shown the pictorial view of the login steps for the password character \textit{A}, 
when index of this character is used for the very first time in an authentication session by $\mathcal{M}$.\\

\textbf{Bypassing the attack in the first session:} Figure \ref{fslc} shows that the original password character of $\mathcal{H}$ always belongs to a group comprising of $\{$A, D, G, J, L, d, e, f, l, o, v, x, y, 0, 6, 8$\}$. These $16$ elements hold the same color and therefore color responses by $\mathcal{H}$ obscure $\mathcal{A}$ among init($\mathcal{X}$) $=16$ possible candidates after recording the first session.

\textbf{Leaked information:} Leaked information after capturing the first session become; $\mathcal{S}_1$ = $\{$A, D, G, J, L, d, e, f, l, o, v, x, y, 0, 6, 8$\}$.

\textbf{Database update:} Including the original password character \textit{A}, ICIP stores all these $16$ elements in the database .\\

\begin{figure}[t!] 
\minipage{0.32\textwidth}
\includegraphics[width=0.7\linewidth]{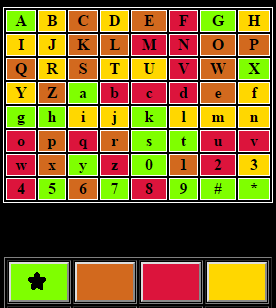}
\subcaption{Round 1: $\mathcal{R}$ is Green} \label{fig:a}
\endminipage\hfill
\minipage{0.32\textwidth}
\includegraphics[width=0.7\linewidth]{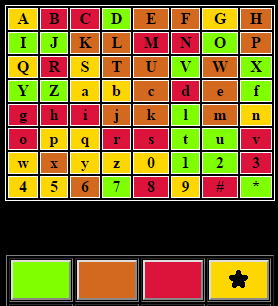}
\subcaption{Round 2: $\mathcal{R}$ is Yellow} \label{fig:b}
\endminipage\hfill
\minipage{0.32\textwidth}
\includegraphics[width=0.7\linewidth]{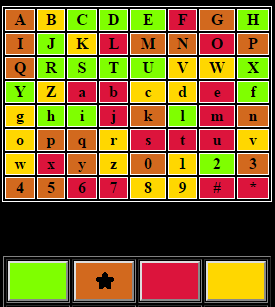}
\subcaption{Round 3: $\mathcal{R}$ is Orange} \label{fig:c}
\endminipage

\minipage{0.32\textwidth}
\includegraphics[width=0.7\linewidth]{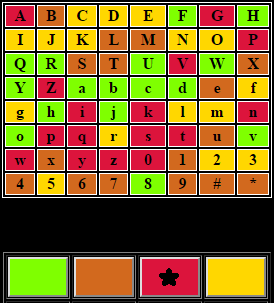}
\subcaption{Round 4: $\mathcal{R}$ is Red} \label{fig:d}
\endminipage\hfill
\minipage{0.32\textwidth}
\includegraphics[width=0.7\linewidth]{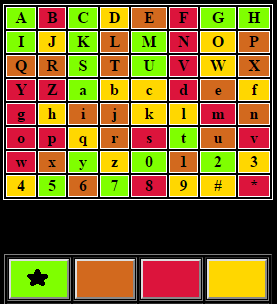}
\subcaption{Round 5: $\mathcal{R}$ is Green} \label{fig:e}
\endminipage\hfill
\minipage{0.32\textwidth}
\includegraphics[width=0.7\linewidth]{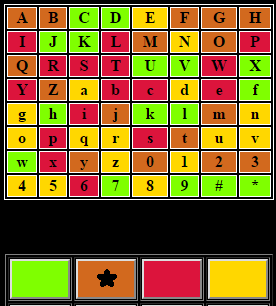}
\subcaption{Round 6: $\mathcal{R}$ is Orange} \label{fig:f}
\endminipage

\caption{\textbf{Login example in ICIP:} Color responses submitted by $\mathcal{H}$ for the password character \textit{A} in each round of the \textbf{second authentication session}. The subgroups are $\{$A, G, y, 0$\}$, $\{$D, J, f, l$\}$, $\{$d, o, v, 8$\}$ and $\{$L, e, x, 6$\}$. In each round, each subgroup gets a unique color. The color button hit by $\mathcal{H}$ in each round is marked by  a ``$\mathlarger{\mathlarger{*}}$" symbol.}
  \label{sslc} 
\end{figure}

\textbf{Bypassing the attack in the second session:} While password character \textit{A} is being used for the second time, $16$ elements from the database are equally distributed among $4$ subgroups in a random manner. Therefore each subgroup contains $\mathcal{Y} =4$ (derived as $\mathcal{Y} = \mathcal{X}/sf$) elements in it. In Figure \ref{sslc}, we present the user authentication for the second session.  The randomly formed subgroups by $\mathcal{M}$ here are $\{$A, G, y, 0$\}$, $\{$D, J, f, l$\}$, $\{$d, o, v, 8$\}$ and $\{$L, e, x, 6$\}$. In each round of this authentication session, each subgroup gets a unique color.

\textbf{Leaked information:} It is quite understandable that as each subgroup in the second session contains $4$ elements, thus, after recording  $\mathcal{H}'$s response in the second session, $\mathcal{A}$ will remain confuse among $4$ possible candidates. Therefore leaked information, $\mathcal{S}_1 \cap \mathcal{S}_2$ yields to $\{$A, G, y, 0$\}$.

\textbf{Probability of threat detection:} If $\mathcal{A}$ does not get a chance to record any further login session after recording the first authentication session, then with the $16$ probable elements from the previous session, $\mathcal{A}$ may try to impersonate the genuine $\mathcal{H}$ in the second session. From the visual interface,  $\mathcal{A}$ first recognizes each subgroup on the grid. But identifying the subgroups only is not sufficient as $\mathcal{A}$ does not know which of them is containing the original password character. As each subgroup holds a unique color therefore, generated sequence of responses for each of them will be different. Thus, if $\mathcal{A}$ selects a wrong subgroup, not holding the original password character, then from the received $\mathcal{R}$, ICIP can detect the recording attack with a probability of $3/4$. 

\textbf{Database update:} After successful login by $\mathcal{H}$ in the second session, $\mathcal{M}$ replaces previously stored $16$ elements by the group of $4$ elements; A, G, y and $0$.\\

\begin{figure}[t!] 
\minipage{0.32\textwidth}
\includegraphics[width=0.8\linewidth]{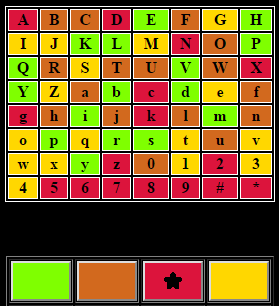}
\subcaption{Round 1: $\mathcal{R}$ is Red} \label{fig:a}
\endminipage\hfill
\minipage{0.32\textwidth}
\includegraphics[width=0.8\linewidth]{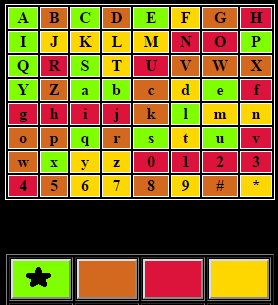}
\subcaption{Round 2: $\mathcal{R}$ is Green} \label{fig:b}
\endminipage\hfill
\minipage{0.32\textwidth}
\includegraphics[width=0.8\linewidth]{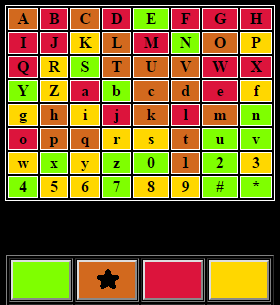}
\subcaption{Round 3: $\mathcal{R}$ is Orange} \label{fig:c}
\endminipage

\minipage{0.32\textwidth}
\includegraphics[width=0.8\linewidth]{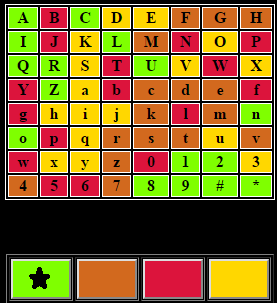}
\subcaption{Round 4: $\mathcal{R}$ is Green} \label{fig:d}
\endminipage\hfill
\minipage{0.32\textwidth}
\includegraphics[width=0.8\linewidth]{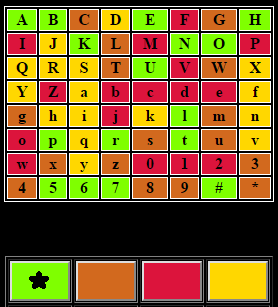}
\subcaption{Round 5: $\mathcal{R}$ is Green} \label{fig:e}
\endminipage\hfill
\minipage{0.32\textwidth}
\includegraphics[width=0.8\linewidth]{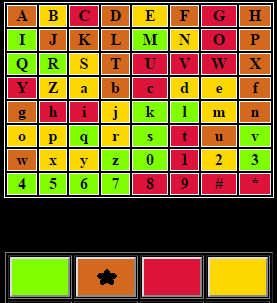}
\subcaption{Round 6: $\mathcal{R}$ is Orange} \label{fig:f}
\endminipage

\caption{\textbf{Login example in ICIP:} Color responses submitted by $\mathcal{H}$ for the password character \textit{A} in each round of the \textbf{third authentication session}. The formed subgroups here are $\{$A$\}$, $\{$G$\}$, $\{$y$\}$ and $\{$0$\}$. In each round, each subgroup gets a unique color. The color button hit by $\mathcal{H}$ in each round is marked by  a ``$\mathlarger{\mathlarger{*}}$" symbol.}
  \label{tslc} 
\end{figure}

\textbf{Bypassing the attack in the third session:} In this session, ICIP generates $4$ subgroups and each of them contains a single element from the set $\{$A, G, y, 0$\}$; where $\mathcal{Y} = 1$, $\mathcal{X} =4$ and \textit{sf} $=4$. In Figure \ref{tslc}, we have shown the login steps that $\mathcal{H}$ follows in the third session. Here also, each subgroup gets a unique color.

\textbf{Leaked information:} At the beginning of third authentication session, $\mathcal{A}$, who already has recorded footage from the previous two authentication sessions, remains confused among the $4$ elements. After recording the submitted $\mathcal{R}$ in the first round only, $\mathcal{A}$ can deterministically extract the original password character as each of those $4$ elements (create confusion initially) holds a unique color. Hence $\mathcal{S}_1 \cap \mathcal{S}_2 \cap \mathcal{S}_3$ reveals $\{$A$\}$. Therefore leakage control strategy can guard a original password character upto $3$ authentication sessions.

\textbf{Probability of threat detection:} Before recording the third session, derived elements from the first two sessions creates confusion among $4$ possibilities in $\mathcal{A}'$s mind. Therefore, if $\mathcal{A}$ tries to login using the recorded footages from the previous two authentication sessions then chances of threat detection still remains as $3/4$.

\textbf{Database update:} After successful login by $\mathcal{H}$ in the third session, $\mathcal{M}$ replaces the previously saved four characters by the original password character.\\

\textbf{Note 2: Strength of ICIP $-$} By virtue of its design, for $\mathcal{P}_2$ of length $4$, ICIP can defeat $\mathcal{A}$ for $3 \times 4 = 12$ authentication sessions. Also, before its utilization in second/third authentication session, each character in $\mathcal{P}_2$ can detect the threat with the probability $3/4$.

\section{Password Management Policy} 
\label{storage}
As mentioned earlier, proposed TPP here includes two passwords ($\mathcal{P}_1$ and $\mathcal{P}_2$) to raise the security bar at both the client and server sides (see detail in Section \ref{sec-ana}). To reinforce the security at server side, $\mathcal{M}$ stores $\mathcal{P}_1$ and $\mathcal{P}_2$ in two different servers. Though $\mathcal{P}_1$ can be stored by following the conventional password storage mechanism, but $\mathcal{P}_2$ needs to store some additional information to fit into the security features of ICIP and, next we will elaborate on this.  

All the recording attack resilient methodologies maintain password in the plaintext format \citep{core-paper} and proposed ICIP is no exception of that. In addition to the username and $\mathcal{P}_2$ (in plaintext), ICIP maintains the group/subgroup information in the database. This is shown in Table \ref{database}.

\begin{table}[!ht]
\centering
\caption{Maintained database by ICIP for username \textit{alice} and password \textit{S7Ay}.}
\begin{tabular}{c c c c c c}
\hline
$\mathcal{H}$ & $\mathcal{P}$ & group-1 & group-2 & group-3 & group-4\\
\hline
alice & S7Ay & S & null  & ADGJLdef & GRy\#\\

      &      &   &       & lovxy068 &   \\
\hline
\end{tabular}
\label{database}
\end{table} 

In the database, group-$i$ can be related to the password character at $i^{\text{th}}$ index. The database shows that first character has already been used for three times and hence index value $1$ must not be used as a challenge for any further authentication session. As a challenge, index values $3$ and $4$ can be used for two and one more authentication sessions, respectively. The database also infers that index value $2$ has never been used by $\mathcal{M}$ and therefore this index value can be used for $3$ times for authenticating $\mathcal{H}$. Proposed scheme will notify $\mathcal{H}$ to  change her password ($\mathcal{P}_2$) once $\mathcal{M}$ makes use of all the index values for $3$ times. Table \ref{database} shows that ICIP incurs some storage overhead.

Many password systems, particularly for government and industry users, store $\mathcal{H}'$s old password $-$ usually
the last 10, as stipulated in, e.g., \citep{report-last10password}. When $\mathcal{H}$ changes her $\mathcal{P}$, she is prohibited in such systems from reusing any stored ones. Many recent literatures argue that instead of storing old $\mathcal{P}$ on a per-user basis, it would be better to record previously used $\mathcal{P}$ across the full user population \citep{honeyword-juels}. A newly created password that conflicts with any password in this list may then be rejected. Instead of storing password in the password-list explicitly, $\mathcal{M}$ may use a Bloom filter \citep{bloom-filter} which would not reveal these passwords directly and sufficiently reduces the storage cost. In this way, the storage overhead of ICIP may be compensated.

\section{Security Analysis}
\label{sec-ana}
Proposed TPP makes use of two passwords $\mathcal{P}_1$ and $\mathcal{P}_2$ for strengthening the security standard from many aspects. This section first captures the $\mathcal{A}'$s activity at server side and illustrates how our scheme performs there to defeat the adversary. Then the focus is shifted towards the client side threat and the defense mechanisms adopted by the TPP for protecting the users' credentials.

\subsection{Managing the server side threat}
\label{sec-ana-01}
The systems $-$ maintaining the login credential of the clients in a single server, often suffer from password leakage once $\mathcal{A}$ gets access to this server. Though advanced techniques like \textit{bcrypt} \citep{bcrypt} provide some relief, but they were never designed to save the passwords against efficient password cracking attacks, even if the passwords are maintained in the hashed format \citep{ma}. Researchers recently have shown that switching towards the distributed environment for storing the passwords can greatly help to address this kind of threat as the distributed framework is much harder to compromise as a whole \citep{sauth}. Under the distributed framework, $\mathcal{M}$ uses more than one server to store the complete password information and as a result, one compromised server does not make $\mathcal{A}$ eligible for impersonating the genuine $\mathcal{H}$. 

Using the proposed TPP, as $\mathcal{H}$ uses two passwords $\mathcal{P}_1$ and $\mathcal{P}_2$ and, $\mathcal{M}$ stores these passwords in two different servers (also discussed in Section \ref{storage}), thus it makes $\mathcal{A}'$s job far more difficult to breach the security at the server side.

\subsection{Managing the client side threat}
\label{sec-ana-02}
Unlike the considered unidirectional threat at server side, $\mathcal{A}$ can perform different kind of attacks for computing $\mathcal{H}'$s secret at client side. The threats at client side primarily include
\begin{itemize}
\item Recoding attack.
\item Guessing attack.
\item Random key submission
\end{itemize}
As proposed ICIP here is a part of login routine (mode of entering $\mathcal{R}$ based on $\mathcal{P}_2$) and an influence from the \textit{honeyword based authentication technique} (HBAT), thus, the security analysis also incorporates HBAT security features like security against DoS attack, multiple system vulnerability and flatness.  Next, we will show how proposed methodology provides security against these attack scenarios.\\

\subsubsection{Evaluating the security standard against the primary threat models}

\textbf{Recording attack:} We already have shown that against recording attack, ICIP can deterministically prevent a password character upto the first round of the third authentication session. So for each authentication session of $6$ rounds, this attack can be prevented upto $13^{\text{th}}$ round and thus, this scheme can save a password character upto  $\lceil$ 13/6 $\rceil$ = $3$ authentication sessions. Therefore, though $\mathcal{P}_1$ immediately gets compromised under the recording attack in a single session, but for  $\mathcal{P}_2$ of length $4$, TPP can defeat $\mathcal{A}$  for $4 \times 3$ = $12$ authentication sessions against this attack.

While Section \ref{intro} shows that most of the methods consider the practical value of $r_{\text{max}}$ as $3$, proposed method here provides security far beyond that limit by setting $r_{\text{max}}$ as $12$.\\

\textbf{Guessing attack:} Let Pr$_{\text{guess}}$[$\mathcal{P}_1$] and Pr$_{\text{guess}}$[$\mathcal{P}_2$] be the probabilities of guessing $\mathcal{P}_1$ and $\mathcal{P}_2$, respectively. Therefore, overall success probability of $\mathcal{A}$ to break the security of TPP under this attack can be evaluated as Pr$_{\text{guess}}$[$\mathcal{P}_1$] $\times$ Pr$_{\text{guess}}$[$\mathcal{P}_2$]. Next we determine the values of Pr$_{\text{guess}}$[$\mathcal{P}_1$] and Pr$_{\text{guess}}$[$\mathcal{P}_2$].

\begin{itemize}

\item \textbf{Evaluating Pr$_{\text{guess}}$[$\mathcal{P}_1$]:} One may find $95$ printable characters on a standard keyboard. Some of the recent studies shows that $\mathcal{H}$ generally prefer to choose password of the length around $6$ \citep{dasTangled}. Therefore for the password of that length, the default value of Pr$_{\text{guess}}$[$\mathcal{P}_1$] can be evaluated as $95^{-6}$ or $1.36 \times 10^{-12}$.\\

\item \textbf{Evaluating Pr$_{\text{guess}}$[$\mathcal{P}_2$]:} Along with guessing $\mathcal{P}_1$, $\mathcal{A}$ need to guess a single password character from the set of $64$ elements to pass through an authentication session and thus, initially it may seem that Pr$_{\text{guess}}$[$\mathcal{P}_2$] yields to $1/64$. But $\mathcal{A}$ may follow a smarter strategy for guessing a single character of $\mathcal{P}_2$ which again can be divided in $3$ scenarios

\begin{itemize}
\item \textbf{Case 1: Guessing a character in $\mathcal{P}_2$ that is used for the first time $-$} As the password character here always belongs to a group of $8$ elements, thus identifying that group only will make $\mathcal{A}'$s job done and the corresponding probability of that will be $8/64$ or $0.125$.

\item \textbf{Case 2: Guessing a character in $\mathcal{P}_2$ that is used for the second time $-$} As the password character here always belongs to a group of $4$ elements, thus identifying that group only will make $\mathcal{A}'$s job done and the corresponding probability of that will be $4/64$ or $0.0625$.

\item \textbf{Case 3: Guessing a character in $\mathcal{P}_2$ that is used for the third time $-$} As the password character here always belongs to a singleton set, thus identifying that set only will make $\mathcal{A}'$s job done and corresponding probability of that will be $1/64$ or $0.0157$.

\end{itemize} 
\end{itemize}

Thus aforementioned discussion infers that likelihood of $\mathcal{A}$ being successful in performing the guessing attack can achieve maximum value $0.17 \times 10^{-12}$. On the other hand, the minimum probability for the same yields to $0.021 \times 10^{-12}$. Therefore, proposed TPP here maintains a strong defense line to refrain $\mathcal{A}$ in performing the guessing attack.\\

\textbf{Random key submission (RKS):} To perform this attack, instead of targeting the actual password, $\mathcal{A}$ randomly presses the response keys to pass through an authentication session. Thus, while the probability of hitting the same key sequences producing $\mathcal{P}_1$ becomes $95^{-6}$ (denoted as Pr$_{\text{RKS}}$[$\mathcal{P}_1$]), the probability of shattering the second line of defense values in $4^{-6}$ (denoted as Pr$_{\text{RKS}}$[$\mathcal{P}_2$]), as the login interface of ICIP includes $4$ response keys only. Hence, the overall probability to break the system under this attack can be represented as Pr$_{\text{RKS}}$[$\mathcal{P}_1$] $\times$ Pr$_{\text{RKS}}$[$\mathcal{P}_2$] which yields to $3.32 \times 10^{-16}$. \\

\subsubsection{Evaluating the security standard against the HBAT threat models}

As discussed earlier, proposed ICIP here is an influence from HBAT. There are three well defined security parameters related to any HBAT \citep{honeyword-juels}. Next we will discuss on how ICIP performs against these attack scenarios.\\

\textbf{DoS attack:} Knowing the original password, if $\mathcal{A}$ can guess a \textit{honeyword} then she may intentionally submit that to mount DoS attack in order to block $\mathcal{H}'$s account \citep{honeyword-juels}.  Let us assume that $\mathcal{A}$ successfully records the first authentication session. Now, if the same password character of $\mathcal{P}_2$ is being used in any of the subsequent sessions and, provided responses by $\mathcal{A}$ in that session justify a group not containing the original password character then genuine user's account may get blocked. Therefore, mounting DoS attack becomes easy after $\mathcal{A}$ records the first session. But as there is really no such legitimate user, $\mathcal{A}'$s attempt is reliably detected when this occurs.\\

\textbf{Multiple system vulnerability:} This is a less practical attack scenario as it depends on few strong assumptions \citep{erguler-first}. Consider a scenario where $\mathcal{H}$ has used the same character of $\mathcal{P}_2$ for login in two different systems. If $\mathcal{A}$ remains active for both these authentication sessions, then she derives two different groups of $16$ elements where each of these holds the original password character. As intersection between these groups reduces the probable candidates, thus the security of ICIP is affected under this situation (ref. to appendix B).

It is important to note here that multiple system vulnerability can threat the security of any recording attack resilient methodology as the adversary gets a chance to record the same user information for more number of authentication sessions.\\  

\textbf{Flatness:}  In HBAT, from the list of probable candidates, if $\mathcal{A}$ cannot differentiate between the original password and \textit{honeywords} then the probability of detecting the attack becomes high. After successfully recording the first authentication session of $\mathcal{H}$, ICIP still confuses $\mathcal{A}$ in identifying the subgroup representing the original secret in the second/third authentication session. Thus ICIP well satisfies the flatness criterion.\\

\noindent\fbox{%
    \parbox{1.0\textwidth}{%
 \textbf{Discussion 5:} \textbf{Some remarks on the security standard of the proposed model}$-$ Proposed TPP here successfully handles the recording attack as $\mathcal{H}$ may use the same login credentials for $12$ authentication sessions. Due to large password space of $95$ printable characters, $\mathcal{P}_1$ leads this scheme towards providing robust security against both the guessing attack and random key submission attack. Also, along with the threat prevention, proposed scheme has a novel property of threat detection; which ``lights two candles with one flame".

It is also meaningful to assume that with the knowledge of probable password characters, $\mathcal{A}$ will always try to identify the correct subgroup rather choosing a wrong one. Therefore $\mathcal{A}'$s mentality of mounting DoS attack seems to be little unrealistic in this context. Launching the recording attack by the same $\mathcal{A}$ at different places by targeting a particular $\mathcal{H}$  also looks way to practical as suggested in \citep{erguler-first} (also, this will degrade security of any human identification protocol).  Thus, by achieving the complete flatness, proposed scheme here not only well satisfies almost all the realistic security aspects at the client side, but also, it manages the threat of password leakage at the server side. Thus, by handling both the client and server side threats efficiently, proposed idea here ensures the ``security beyond the conventional limit".       
    }
   }

\section{Usability analysis}
\label{use-ana}
In TPP, the submission of $\mathcal{P}_1$ requires no manipulation from $\mathcal{H}'$s end and thus, it provides the same usability standard as of the legacy password protocol. But in order to generate $\mathcal{R}$ with reference to $\mathcal{C}$, $\mathcal{H}$ needs to process $\mathcal{P}_2$ accordingly and this plays a major role in determining the overall usability standard of the TPP. Therefore, (until specified explicitly) this section mainly focuses on the usability standard provided by the ICIP.

 We determine the usability standard of the ICIP both from the theoretical and experimental point of views. As discussed in \citep{core-paper}, the theoretical analysis here is independent of any particular user set and hence the outcome remains static under any situation. The usability standard also includes the HBAT usability parameters which are (a) system interference, (b) stress-on-memorability and (c) typo safety.

\subsection{Theoretical analysis:} This framework is mainly driven by two components $-$ \textit{Cognitive Workload} ($\mathbb{CW}$) and \textit{Memory Demand} ($\mathbb{MD}$). $\mathbb{CW}$ influences the login time and is measured against total reaction time (in seconds) required by the atomic cognitive operations. There are four well defined atomic cognitive operations associated with a human identification protocol, and these are 
\begin{itemize}
\item  (Single/Parallel) Recognition \citep{yan-28}
\item  (Free/Cued) Recall \citep{yan-23}
\item  (Single-target/Multi-target) Visual Search \citep{yan-33}
\item  Simple Cognitive Arithmetic \citep{yan-6}
\end{itemize}

\textbf{Cognitive overload:} From the visual challenge, sending the index value, $\mathcal{H}$ first performs a cued recall to perceive her password character. The reaction time for cued recall ($\mathbb{CR}$) can be obtained through (0.3694+0.0383 $\times$ \textit{g} $\times$ $\psi$), where \textit{g} (default value is $1$ here) denotes number of elements $\mathcal{H}$ requires to remember for performing the login in a round \citep{yan-23} \citep{yan-7}. $\psi$ is the ratio between the cued recall and single item recognition, and the default value of this is $1.969$ \citep{yan-23}. For $\mathcal{P}_2$ of length $4$, $\mathcal{H}$ has to perform cued recall once at the beginning of a 
authentication session consisting of $6$ rounds. Thus, average cued recall time for each round can be calculated as ($\mathbb{CR}$/6)$\times$ 4 which yields $\alpha_1 = 0.448 \times \frac{2}{3}$ or $0.299$.

In order to derive a response, $\mathcal{H}$ requires to search the correct clue from a set of \textit{w} (also refers as window size) varying objects, displaying on the screen. During identification of a single target, the reaction time of $\mathcal{H}$ can be formulated as $\alpha_2 = 0.583+0.0529 \times w$ \citep{yan-34}. By using ICIP, $\mathcal{H}$ needs to identify a color from a single static cell on the grid through out a session. Therefore, $w$ can be assumed as $1$ here which returns the value of $\alpha_2$ as $0.6359$. 

Above discussion infers that ICIP derives $\mathbb{CW}$ as $\alpha_1 + \alpha_2 = 0.9349$ for each round.\\

\textbf{Memory demand:} For $\mathbb{MD}$ operation, the cost of each scheme can be calculated as the ratio between length of  $\mathcal{P}_2$ and $\lambda_{op}$: where $\lambda_{op}$ is the accuracy rate of corresponding memory retrieval operation within a  fixed memorization time. Since recognition is much easier than recall  therefore,  $\lambda_{op}$ becomes $29.6\%$ and $84.8\%$ for recall and recognition, respectively \citep{yan-14}. Thus $\mathbb{MD}$ in ICIP can be calculated as $13.51$, for the length of $\mathcal{P}_2$ as $4$ and $\lambda_{op}$ as $29.6\%$.\\

Finally, an overall score of human power, $\mathbb{HP}$,  can be derived as the product of $\mathbb{CW}$ \textit{for a session} and $\mathbb{MD}$. Thus, $\mathbb{HP}$ for ICIP can be obtained as $0.9349 \times 6 \times 13.51 = 75.78$, which is  significantly less compared to the existing protocols in this direction (see details Section \ref{comp}).

\subsection{Experimental analysis}
To conduct experimental analysis, we took help from $93$ participants (all having correct-to-normal eyesight) and identified this set of participants as $\mathbb{P}^*$. The participants were capable of operating computers and their age varied between $19$ to $37$.  Among them, $18$ participants were identified as skilled $\mathcal{H}$ as they enjoy playing fast video games \citep{kwon2014}. It is important to note here that according to the report in \citep{no-of-user-guide-1}, most of the studies in this area used a test group between $11$ and $25$, while very few used a test group of $50$ or more participants \citep{no-of-user-guide-2}. Therefore, we believe that a set of $93$ participants will provide more base to our experimental study.

The analysis was conducted into two phases $-$ \textit{training phase} and \textit{test phase}. In \textit{training phase}, we first gave the participants a quick motivation behind our work and  demonstrated the working principle of the proposed TPP. For illustrating the model, we used the $10$ Desktops of our institute laboratory. Thereafter, we asked each of them for login by using the predefined login credentials set by us. After each participant performed the login, we gave them an entire day to use the proposed model (in the laboratory) for getting habituated with it. In the \textit{test phase}, we collected the test data for $5$ days and each participant was allowed to login for $3$ times. Obtained login time from skilled $\mathcal{H}$ helped us to determine the accuracy of derived $\mathbb{CW}$ for the ICIP in the previous section.

From $93 \times 3 = 279$ login attempts, below we present some notable information/outcomes of our experiment.

\begin{itemize}
\item The ratio between the skilled and non-skilled participants was $1:4$ (approximately).
\item $16$ times participants failed to input $\mathcal{P}_1$ correctly.
\item $23$ times participants failed in generating the valid responses by using $\mathcal{P}_2$.
\item Overall $28$ times they failed to login using TPP.
\item Particularly for ICIP, the average login time of the skilled participants was recorded as $1.31$ ($7.96/6$) seconds per round, close enough to the measured $\mathbb{CW}$ parameter with an allowable small difference \citep{core-paper}. 
\item The average input submission time during entering $\mathcal{P}_1$ (set as ``anhour") was obtained as $3.2$ seconds considering the successful login attempts only.
\item The average successful secret submission time for all the participants was captured as $15.49$ seconds (for entering both the secret information $\mathcal{P}_1$ and $\mathcal{P}_2$).
\end{itemize}

After conducing the experiment we asked the participants to fill a feedback form to rate our proposed model. The obtained feedback result was found promising and is presented in Table \ref{feedback}.

\begin{table}[!ht]
\centering
\caption{Recorded feedbacks from the $93$ participants}
\begin{tabular}{c c c}
\hline
Choices & Agreed participants & Percentage ($\%$)\\
\hline
Love to use & $19$ & $20.43$ \\
\hline
Easy to use & $26$ & $27.95$\\
\hline
Find usable & $35$ & $37.64$\\
\hline
Bit difficult & $8$ & $8.6$\\
\hline
Extremely difficult & $2$ & $2.15$\\
\hline 
Not sure & $3$ & $3.23$\\
\hline
\end{tabular}
\label{feedback}
\end{table}

\subsection{Evaluating ICIP under the HBAT usability features}

\textbf{System interference:} If a \textit{honeyword} based approach influences the password choice of $\mathcal{H}$ (e.g., forces $\mathcal{H}$ to remember some extra information) then it significantly impacts on users' convenience. As the proposed model does not influence the password choice of $\mathcal{H}$, thus system interference can be considered as negligible.\\.

\textbf{Stress on memorability:} If a \textit{honeyword} based approach influences the password choice then that may put some burden on $\mathcal{H}'$s mind as she may need to remember some additional information. This feature also has severe impact on usability standard and proposed ICIP does not include this too.\\

\textbf{Typo safety:} A \textit{honeyword} generation algorithm is called typo safe if typing mistake of $\mathcal{H}$ rarely matches with any \textit{honeyword}. Let us assume that response sequences generated by two subgroups in ICIP are $< C^{g1}_1-C^{g1}_2...C^{g1}_5-C^{g1}_6>$ and $< C^{g2}_1-C^{g2}_2...C^{g2}_5-C^{g2}_6>$; where $C^{gi}_k$ symbolizes color response corresponding to group $i$ in the $k^{th}$ round. Without loss of generality, if we assume that the secret bit of $\mathcal{P}_2$ is belonging to group $1$ then $\mathcal{H}$ follows the response sequence $< C^{g1}_1-C^{g1}_2...C^{g1}_5-C^{g1}_6>$.  For any authentication round $i$ ($1 \leq i \leq 6$), as $C^{g1}_i$ and $C^{g2}_i$ always differ, thus a few typing mistakes will never yield to $< C^{g2}_1-C^{g2}_2...C^{g2}_5-C^{g2}_6>$. Therefore proposed ICIP is highly typo-safe.\\

\noindent\fbox{%
    \parbox{1.0\textwidth}{
    
\textbf{Discussion 6:} \textbf{Some remarks on the usability standard of the proposed scheme}$-$ Theoretical analysis shows that the second line of defense, ICIP, sets a very low value of  human effort. Experimental analysis suggests that the login time (for entering both $\mathcal{P}_1$ and $\mathcal{P}_2$) can be kept within a range of $9-17$ seconds with an error rate around $10\%$. ICIP also meets with all the usability standards of a HBAT which infer that proposed TPP as a whole is highly usable in practice.
 } 
}

\section{Comparative Analysis}
\label{comp}
For comparing proposed TPP with the existing protocols, we only consider those methods which are usable by atleast $80\%$ of the users (as reported in the respective literatures). In contrast, hard to use  methods, like low complexity CAS \citep{strong-ssa-sp} which demands $30$ objects to be remembered by $\mathcal{H}$ or, HB protocol \citep{strong-ssa-hopper}, requires cryptographic computation from $\mathcal{H}$, have been kept outside of this comparative study.

\begin{table*}[!ht]
\caption{Comparative analysis of methods in terms of security features. Session resiliency against recording attack for S3PAS and CHC protocol are obtained from the analysis reported in \citep{core-paper}. While security of PAS against recording attack is taken from the analysis, made by Li et. al \citep{sec-pas}. Variable s ($\geq 0$) in the above table denotes an integer value. LR denotes number of login rounds in a session. $\circledast$ indicates except the first authentication session.}

\centering
\resizebox{0.9\textwidth}{!}{
\begin{tabular}{c c c c c c c c c}
\hline
Method & Secret        & Total                            & Window               & Password  & Pr[RKS] & LR     & Session & Pr[Threat            \\
       &length ($\ell$)& elements  (\textit{n})           &  size (\textit{w})   & space     &/round   &        & resiliency & detection] \\
\hline

CHC    &  $\textbf{5}$  & $\textbf{112}$  & $\textbf{83}$   & $\textbf{1.341} \times \textbf{10}^\textbf{8}$  & $\textbf{0.22}$ & $\textbf{5}$ &          $\textbf{3}$   & \textbf{0}     \\
\hline
PAS    &  $\textbf{4+2s}$  & \textbf{N/A}   & $\textbf{13}$  & $\textbf{4.225} \times \textbf{10}^\textbf{5}$  & $\textbf{0.25}$ & $\textbf{4}$ & $\textbf{9+s}$ & \textbf{0}\\
\hline
S3PAS  &  $\textbf{4}$  & $\textbf{94}$   & $\textbf{94}$  & $\textbf{7.9}\times \textbf{10}^{\textbf{7}}$  & $\textbf{0.076}$ & $\textbf{4}$ &$\textbf{8}$  & \textbf{0} \\
\hline
TPP   & $\textbf{6+} \textbf{4}$ & $\textbf{95}$ & $\textbf{64}$ & $\textbf{1.2} \times \textbf{10}^{\textbf{19}}$ & $\textbf{1.3} \times {\textbf{10}}^{-\textbf{12}}$ $\times \textbf{0.25}$ & $\textbf{1+6}$ & $\textbf{12}$  & \textbf{0.75} $\circledast$\\
\hline          
\end{tabular}}
\vspace{0.2cm}
\label{sec-comp}
\end{table*}

\begin{table*}[!ht]
\caption{Comparative analysis of methods in terms of usability features. $\mathbb{CW}$ per round, for all the methods (except S3PAS) is obtained from \citep{core-paper}. For S3PAS, $\mathbb{CW}$/round is calculated by following the same direction as mentioned in Section \ref{use-ana}. LR indicates number of login rounds in a session.}

\centering
\resizebox{0.9\textwidth}{!}{
\begin{tabular}{c c c c c c c c}
\hline
Method   & LR & Avg. login time & Avg. login time &  Error rate ($\%$)  & $\mathbb{CW}$/round           & $\mathbb{MD}$  & $\mathbb{HP}$ = $\mathbb{CW}$ $\times$ LR \\ 
         &      & skilled user (sec)  & (sec) all user  &   all user          & (sec)         &     &  $\times \mathbb{MD}$ ($\times 10^2$)\\
\hline
CHC      & \textbf{5} &   $\textbf{56}$       &    $\textbf{65.5}$      &   $\textbf{17.1}$    & $\textbf{9.326}$ & $\textbf{16.89}$  &  $\textbf{7.87}$       \\
\hline
PAS      & \textbf{4} &   $\textbf{33.44}$       &    $\textbf{41.52}$     &   $\textbf{15.2}$    & $\textbf{6.837}$            & $\textbf{13.51}$  &  $\textbf{3.69}$        \\
\hline
S3PAS    & \textbf{4} &   $\textbf{36.56}$       &    $\textbf{50.8}$      &   $\textbf{15.7}$    & $\textbf{10.597}$           & $\textbf{13.51}$  &  $\textbf{5.55}$        \\
\hline
TPP     & \textbf{1+6} &   $\textbf{1.3 + 7.86}$       &    $\textbf{3.2 + 12.29}$      &   $\textbf{10.04}$     & $\textbf{0.9349}$           & $\textbf{20.27 + 13.51}$  &  $\textbf{1.45 + 0.7578}$           \\
\hline 
\end{tabular}}
\vspace{0.2cm}
\label{use-comp}
\end{table*}

Because of involvement of human intelligence factor, providing good usability standard is a must criterion of an adoptable human identification protocol. We compare TPP with three existing usable protocols (also certified by the authors in \citep{core-paper}), Convex-Hull-Click (CHC) \citep{chc}, PAS \citep{strong-ssa-pas} and S3PAS \citep{strong-ssa-s3pas}. The experimental data (for usability analysis) were obtained from participants belonging to $\mathbb{P}^+$ set: where $\mathbb{P}^+ \subseteq \mathbb{P}^*$ and as of $\mathbb{P}^*$, $\mathbb{P}^+$ maintains a steady ratio between the skilled and non-skilled $\mathcal{H}$ around $1:4$ for obtaining the unbiased results. Table \ref{sec-comp} and Table \ref{use-comp} show comparative study of the methods from the security and usability perspectives, respectively. 

It is important to note here that submission of $\mathcal{P}_1$ does not demand any significant $\mathbb{CW}$, but requires some $\mathbb{MD}$. Therefore, for the default length of $\mathcal{P}_1$ as $6$, $\mathbb{MD}$ can be calculated as $20.27$.

Along with Discussion $5$ and Discussion $6$ in the previous section, the comparative analysis shows that proposed scheme stands strong in terms of fulfilling all the security aspects. Also, with a remarkable property of threat detection, TPP attains the highest usability standard among all.

\section{Conclusion}
\label{conclusion}
In this paper we have tried to break the chain of ``Drawback-SmallImprovement-Drawback-SmallImprovement" in the research domain of human identification protocol for addressing the threat of recording attack. Propose mechanism here not only reduces the human effort at a large scale, but also offers few security features that are missing in the existing state of arts. The usage of two passwords contributes in enhancing the security standard from many aspects like managing both the server and client side threats. To the best of our believe, we, for the first time introduce the idea of \textit{honeyword} into recording attack resilient unaided human identification protocol design. Exploring the concept of \textit{honeyword} for threat detection, ensuring the security at both the (client and server) ends and extremely simple login procedure make the proposed idea deployable in practice.

\section{References}

\balance

\appendix
\normalsize
\section{. Honeyword based authentication technique (HBAT)}
Once an adversary gets an access to a system, she may obtain the password file where login credential of a registered user is maintained. After retrieving the password information, adversary may impersonate the genuine user.  To resist this kind of threat, system does the following trick. System first creates $\mathcal{K}-1$ ($> 0$) fake passwords and stores those in  password file along with the original password. Thus for username \textit{alex} and password as \textit{infosec}, system may maintain the following list of passwords for $\mathcal{K} = 5$.

\begin{center}
\begin{tabular}{c c c c c}
trustin & privsec & \textbf{infosec} & intrust & secpriv
\end{tabular}  
\end{center}

Together with username, the original password index (here $3$) is maintained in a different file in a different server, known as \textit{honeyChecker}. Thus, compromised password file here obfuscates the adversary among $\mathcal{K}$ possible options. Choosing a wrong password (\textit{honeyword}), instead of original one, will send an incorrect index value to the \textit{honeyChecker} and through this the attack can be detected. In a nutshell, by creating a distributed security environment, HBAT can detect the attack on the password file with a probability of $\frac{\mathcal{K}-1}{\mathcal{K}}$. 

\section{. Evaluating Security against multiple system vulnerability (MSV)}
To show the effect of multiple system vulnerability (MSV) on ICIP, we consider a base case (close to practical) where an adversary is active over $2$ different systems. Now if a user uses the same password character for login in those two different systems then following are the probable cases which may threat security of a password character under MSV.

\textbf{\textit{After recording the first authentication sessions}} of a user during login into both the systems, an adversary derives two different lists probable candidates. Both these lists contain the original password character. Finding more than one common elements between these two lists restricts $\mathcal{A}$ to derive the original secret deterministically. Hence, the success probability of $\mathcal{A}$ becomes $\mathbb{P}_{\text{Disclosure}} = 0.007$; where value of $\mathbb{P}_{\text{Disclosure}}$ has been derived from the Equation \ref{eqnMSV} by taking \textit{m} as $15$.   

Adversary may \textbf{\textit{record second authentication sessions}}, while the same password character is used for login purpose into both of those systems. To avoid the effect of MSV, adversary must find atleast two common characters between the groups containing the original password character. For ICIP, probability of this can be calculated as 

\begin{equation}
\text{Pr} = \bigg[1-\prod_{k=1}^{15}\dfrac{63-k}{64-k}\bigg]\times \dfrac{1}{4}
\end{equation} 
which yields to 0.06. Therefore, effect of MSV may proven to be catastrophic in this scenario.

In anyway, as adversary gets to know the exact secret while a password character is used for the third time, hence it wouldn't be much meaningful to derive the effect of MSV after \textbf{\textit{recording the third authentication sessions}}.
\balance

\end{document}